\documentclass{raa}

\usepackage{graphicx,times}             

\begin{document}

   \title{The surge-like eruption of a miniature filament
}

   \volnopage{Vol.0 (200x) No.0, 000--000}      
   \setcounter{page}{1}          

    \author{Jia-Yan Yang
      \inst{1,2}
    \and Yun-Chun Jiang
      \inst{1}
    \and Dan Yang
    \inst{1,3}
    \and Yi Bi
    \inst{1,3}
    \and Bo Yang
    \inst{1,3}
   \and Rui-Sheng Zheng
   \inst{1,3}
   \and Jun-Chao Hong
   \inst{1,3}
   }

   \institute{National Astronomical Observatories/Yunnan Observatory, Chinese Academy of Sciences,
             Kunming 650011, China; {\it yjy@ynao.ac.cn}\\
        \and
             Key Laboratory for the Structure and Evolution of Celestial Objects,
            Chinese Academy of Sciences, Kunming 650011, China\\
        \and
             Graduate School of Chinese Academy of Sciences, Beijing 100049, China\\
   }

   \date{Received~~2011 month day; accepted~~2011~~month day}

\abstract{ We report on the rare eruption of a miniature H$\alpha$
filament that took a surge form. The filament first underwent a full
development within 46 minutes and then began to erupt 9 minutes
later, followed by a compact, impulsive X-ray class M2.2 flare with
a two-ribbon nature only at the early eruption phase. During the
eruption, its top rose, whereas the two legs remained rooted in the
chromosphere and swelled little perpendicular to the rising
direction. This led to a surge-like eruption with a narrow angular
extent. Similar to the recent observations for standard and blowout
X-ray jets by Moore et al., we thus define it as a ``blowout
H$\alpha$ surge''. Furthermore, our observations showed that the
eruption was associated with (1) a coronal mass ejection guided by a
preexisting streamer, (2) abrupt, significant, and persistent
changes in the photospheric magnetic field around the filament, and
(3) sudden disappearance of a small pore. These observations thus
provide evidence that blowout surge is a small-scale version of
large-scale filament eruption in many aspects. Our observations
further suggest that at least part of H$\alpha$ surges belong to
blowout-type ones, and exact distinction between standard and
blowout H$\alpha$ surges is important in understanding their
different origins and associated eruptive phenomena.
   }

   \authorrunning{J. Y. Yang et al.}            
   \titlerunning{The surge-like eruption of a miniature filament}  

   \maketitle

%
%
\section{Introduction}           
\label{sect:intro}

Solar filaments show some common characteristics in a broad spectrum
of scales. Along the polarity inversion zone between adjacent
opposite-polarity photospheric field on the quiet Sun, miniature
filaments are the small-scale analog to large-scale ones. Hermans \&
Martin (1986) concluded that their eruptions appear
to be the counterparts of large-scale filament eruptions, which are
usually associated with two-ribbon flares and coronal mass ejections
(CMEs). Wang et al. (2000) have shown that miniature
filaments have a mean lifetime of only 50 minutes, and their
eruptions can take varying forms but almost all are accompanied by
tiny flares, with spatial patterns very similar to
two-ribbon/multiribbon flares in large-scale filament eruptions.
Sakajiri et al. (2004) and Ren et al. (2008)
also suggested that small- and large-scale filament
eruptions have common properties. Therefore, it is reasonable to
expect that some small-scale filament eruptions can also relate to
CMEs (Innes et al. 2009; Schrijver 2010) and should be explained
in the framework of the
``standard model'' for solar eruptions, in which overlying arcade
erupts, producing a flare along with an ejecting filament. However,
Wang et al. (2000) showed that most miniature
filaments are erupted toward nearby strong network elements, meaning
that perhaps most of their mass is transported to other magnetic
structures rather than ejected into corona. Therefore, questions are
naturally raised: is the mass transport along pre-existing magnetic
force lines? Can small-scale filament eruptions be strong enough to
escape into the heliosphere? Detailed observations of small-scale
filament eruptions could help to answer these questions.

Strongly related dynamical phenomena that can be closely associated
with CMEs are plasma ejections following open or far-reaching field
lines, which show up as surges in H$\alpha$ and jets in EUV and
X-ray. H$\alpha$ surges are straight or slightly curved mass
ejections stretching out and away from small flares at the
footpoints in the chromosphere into coronal heights (Roy 1973;
Bruzek \& Durrant 1977), and it is believed that they are produced
by reconnection between emerging flux and ambient open or
far-reaching magnetic structures (Schmieder, Van Driel-Gesztelyi, \&
Freeland 1995; Canfield et al. 1996; Jiang et al. 2007; Madjarska et
al. 2007; Chifor et al. 2008; Moreno-Insertis et al. 2008;
Patsourakos et al. 2008; Pariat, Antiochos, \& DeVore 2010; Jiang et
al. 2011; Yang et al. 2011). Therefore, surges have quite different
physical processes and scenarios from those of small filament
eruptions. However, some observations have showed that both of them
can be related to the same kind of eruptive phenomena, specially
``narrow CMEs'' with angular widths of about 15\degr\ or less and
impulsive solar energetic particle (SEP) events. Gilbert et al.
(2001) concluded that 15 narrow CMEs originate in regions of closed
magnetic fields since they have a high association with filaments,
but Dobrzycka et al. (2003) cannot definitively determine whether
they origin from jets or filament eruptions. Liu (2008) showed that
large jetlike/diffuse H$\alpha$ surges are associated with
jetlike/wide-angle CMEs. It is noted that, however, their H$\alpha$
data might have inadequate spatial resolution to tell us whether the
surge bases include tiny filaments or not. Meanwhile, when
white-light jets can be accompanied by either EUV bright points
(Paraschiv et al. 2010) or filament eruptions (Wang \& Sheeley
2002), impulsive SEP events can also be closely associated with EUV
and corresponding white-light jets that involve open field lines
(Wang, Pick, \& Mason 2006), as well as with motions that look like
filament eruptions (Nitta et al. 2006). In order to get a clear
physical picture of these eruptive phenomena, it is clear that a
definitive distinction of their different origins is desired.

On the other hand, some observations indeed presented evidence that
plasma ejections can be physically related with some kinds of
activations or eruptions of small filaments. Chae et al.(1999) first
reported that H$\alpha$ surges, EUV jets, and associated microflares
were preceded by a small filament eruption. Zuccarello et al.(2007)
showed that the destabilization and disappearance of a small
short-lived filament followed two H$\alpha$ surges pouring sequentially
from part of the filament body. Consistent with previous results
(Yamauchi et al. (2005), in particular, Moore  et al. (2010) found that
in the polar coronal holes (CHs) two thirds of X-ray jets are standard
ones that fit the standard reconnection picture for coronal jets, while
other one third are so-called blowout jets in which the jet-base magnetic
arch, often carrying a filament, undergoes a miniature version of
the blowout eruptions that produce major CMEs. Similarly,
Raouafi et al. (2010) showed that coronal jets can erupt from coronal
micro-sigmoids, and Nistic\`o et al. (2009) also found some micro-CME-type
jet events resembling the classical CMEs. Quite recently,
Hong et al. (2011) and Zheng et al. (2011) indeed showed that miniature
filament eruptions can lead to blowout jets and small CMEs.
Therefore, at least part of plasma ejections are originated from
small-scale filament eruptions, and high-resolution observations are
needed to understand the possible relation between them or to
distinguish them from each other. Similar to the cases in
large-scale filament eruptions, in addition, a few examples clearly
suggested that flux emergence and cancelation play an important role
in disturbing small filaments (Hermans \& Martin 1986; Sakajiri et al. 2004;
Ren et al. 2008; Hong et al. 2011). It thus appears that
small filament eruptions have similar
exterior driving agents to that of surges or jets (Yoshimura, Kurokawa,
\& Shine 2003; Jiang et al. 2007). Like the cases occurring in some major
flares (Wang et al. 2002b; Sudol \& Harvey 2005; Petrie \& Sudol 2010;
Wang \& Liu 2010), an abrupt,
significant, and persistent change in photospheric magnetic field
was also found in a CME-associated H$\alpha$ mass ejecta event (Uddin et al.
2004; Li et al. 2005). Therefore, it is probable that such extreme
change could also take place in small-scale filament eruptions, and
detailed magnetic filed observations could help in verifying this
possibility.

On 2001 March 27, a surge-shaped eruption occurred in AR 9401
(N21\degr, E33\degr), and was associated with a compact impulsive
flare, a CME, and especially, a sudden and permanent change of
magnetic field and disappearance of a small pore. High-resolution
H$\alpha$ observations from Big Bear Solar Observatory (BBSO) help
us to find that it was in fact a small-scale filament eruption. Due
to the short filament lifetime, the complete process from its
formation to eruption was observed, hence provided us with the
opportunity to investigate why the eruption took a surge form and
its relationship with the CME and photospheric field activity.


\section{Observations}
\label{sect:Obs}

The eruption was well covered by full-disk H$\alpha$ line-center
observations from BBSO, with a 1-minute image cadence and a pixel
size of  roughly $1''$. The magnetic field configuration of the
eruptive region was examined using full-disk magnetograms and
continuum intensity images from the Michelson Doppler Imager (MDI;
Scherrer et al. 1995) aboard the {\it Solar and Heliospheric
Observatory} ({\it SOHO}). The magnetograms have a 1-minute cadence
and a pixel size of $2^{''}$, while continuum images only have a
96-minute cadence. The MDI data are obtained from nine filtergrams
taken in five different position in the Ni {\sc i} 6768 \AA\
absorption line in right- and left-circularly polarized light, and
the continuum intensity is estimated from a filtergram averaging
narrowband signals from both sides outside the line. The full-disk
BBSO to MDI co-alignment was done by fitting the solar limbs, with
an accuracy of about $2^{''}$.

This eruption was also examined by using of full-disk EUV
observations from the Extreme Ultraviolet Telescope (EIT;
Delaboudini\`{e}re et al. 1995) on {\it SOHO}. The 12-minute cadence 195
\AA\ images were obtained for the study with a pixel resolution of
$2.6^{''}$. To identify the associated CME, we used the C2 and C3
white-light coronagraph data from the Large Angle and Spectrometric
Coronagraphs (LASCO; Brueckner et al. 1995) on {\it SOHO}, as well as
the CME height-time data that is available in the LASCO Web site.
Finally, we used soft X-ray light curves observed by the {\it
Geostationary Operational Environmental Satellite} ({\it GOES}) to
track the time of the flare.

\section{Results}
\label{sect:results}

The surge-like solar eruption was accompanied by a flare of
H$\alpha$ importance 1N and X-ray class M2.2, which {\it GOES}
recorded its start, peak, and end times around 16:25, 16:30, and
16:32 UT, respectively. Fig. 1 presents BBSO H$\alpha$ line-center
images to show the morphological evolution of the event. To aid
matching, an MDI magnetogram is given in the first frame. The
H$\alpha$ flare appeared as a compact one. It started to brighten
slightly prior to the {\it GOES} flare start time, increased in
size, flared through the period of the {\it GOES} flare, quickly
faded away, and finally disappeared after about 16:40 UT. Therefore,
the flare behaved as a compact, impulsive one with a total duration
of only 7 minutes. A striking characteristic of the event is that a
surge-like mass ejection occurred in the course of the flare, which
showed up as extended linear structures with much larger size than
that of the compact flare patch. When the {\it GOES} flare started
(16:25 UT), a single bright structure (indicated by the white arrow)
was first ejected from the flare patch. By the {\it GOES} flare peak
time (16:30 UT), however, the bright structure clearly evolved into
a narrow bright loop (indicated by the white arrow) with two legs,
`1' and `2' (indicated by the two black arrows). The leg 2 quickly
became ambiguous, while the leg 1 largely lengthened towards the
northeastern direction and showed slightly curved triangular shape,
thus remarkably displayed as a surge (Kurokawa \& Kawai 1993). Beginning
at about 16:37 UT after the {\it GOES} flare ended, two dark
components, also labeled as `1' and `2', grew nearly along the same
trajectories of the two bright legs (indicated by the two black
arrows). When the dark 1 gradually grew to occupy the whole path of
the bright 1, the dark 2 underwent a larger development than that of
the bright one. It is possible that the dark components were caused
by the cooling of earlier, brighter components since they were
co-spatial and sheared the same paths. Then the two dark components
gradually faded away and became invisible after about 16:55 UT.

   \begin{figure}
   \centering
   \includegraphics[width=\textwidth, angle=0]{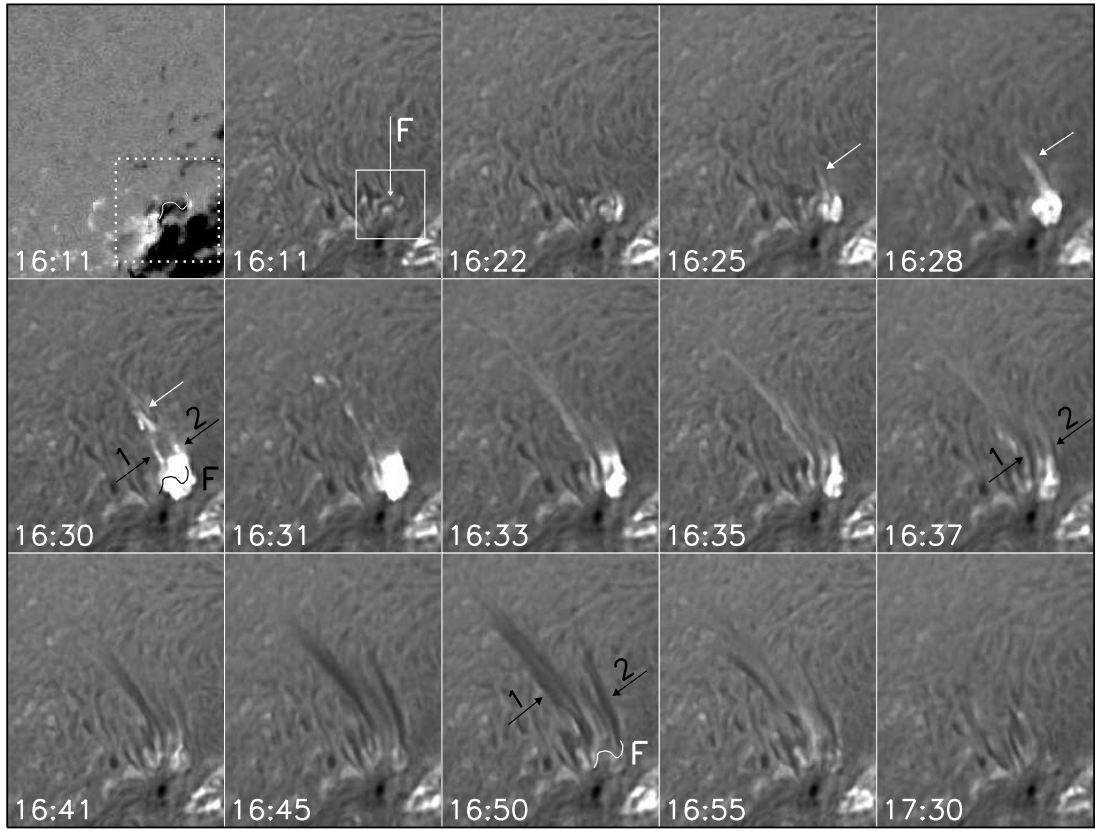}  
   \caption{BBSO H$\alpha$ line-center images showing the evolution of
the event. The first panel is an MDI magnetogram. Accompanied by a
compact flare, the surge-like mass ejection with two leg-shaped
features, `1' and `2', was originated from the eruption of a tiny
filament, `F', at its base. The outlines of F's axis determined from
the 16:11 UT H$\alpha$ image are plotted as solid curves. The
field-of-view (FOV) is $190^{''}$ $\times$ $240^{''}$. The
solid/dashed box indicates the FOV of Fig. 2/Fig. 5.  }
   \label{fig1}
   \end{figure}

Since H$\alpha$ surges can be entirely in emission and dark surges
can be preceded by bright ones (\v{S}vestka, Farnik, \& Tang, 1990),
at first glance,
the above morphological characteristic of the eruption gives us an
impression that this was a surge activity consisting of both bright
and dark components, which occurred one after another, first bright
then dark. Thanks to the high-resolution H$\alpha$ observations,
however, we can remark by chance that there existed a tiny filament,
`F', at the flare site before the eruption (see the 16:11 UT frame
in Fig. 1). Moreover, we found that the legs 1 and 2 were coincided
with those of the erupting F (see the 16:30 and 16:50 UT frames in
Fig. 1). Therefore, it is very likely that the surge-like ejection
was come from the F eruption. This motivates us to further
investigate the F evolution in details. Fortunately, BBSO's
observations covered the key process of the F formation and
eruption. This is shown by the close-up view of F in Fig. 2. When
BBSO's observations started at 15:25 UT, F was seen as a very small
dark feature nearly residing above a magnetic polarity reversal
boundary between adjacent opposite-polarity photospheric field. In
less than 1 hr, it grew toward the west with a curved path along the
polarity reversal boundary (indicated by the white arrows), and by
about 16:11 UT, reached its maximum extent with a length of about
3.86 $\times$ $10^{4}$ km, about two times as long as miniature
filaments with an average projected length of 1.9 $\times$ $10^{4}$
km (Wang et al. 2000). Therefore, F was a somewhat larger miniature
filament. It is of interest to note that at this time it exhibited
an inverse-S shape, consistent with the preferential pattern of
dextral filament in the northern hemisphere (Pevtsov et al. 2003). But
soon afterwards, its westernmost part began to disappear and two
bright patches appeared at its two sides just before the {\it GOES}
flare started (indicated by the black arrows in the 16:20 and 16:24
UT images). This is consistent with Kahler et al. (1988) that filaments
began to erupt before the starting times of associated flares.
Finally, the entire F quickly became invisible along with the area
increase of the flare. Since F did not recover from such
disappearance in the following 2h even when the flare completely
faded away, we can conclude that it really underwent an overall
violent eruption rather than was simply covered and obscured by the
flare emission. It is noted that there were some filament-like dark
features around F that were disturbed or even partially disappeared
during the eruption. By carefully examining the erupting process,
however, it is found that the legs 1 and 2 did not belong to them
but just were the two legs of the erupting F. This is clearly
indicated by the 16:30 and 16:50 UT H$\alpha$ images, where the
outlines of F's axis at 16:11 UT were superposed as dashed lines.
Therefore, the surge-like ejection was simply originated from the F
eruption at its base in a narrow way: when its two ends were fixed
in the chromosphere, its top quickly lifted up but its axis only had
a little swelling or expanding perpendicular to the rising direction
(also see Fig. 1). This mean that they were the same phenomenon,
i.e., a filament eruption taking a surge form (Nistic\`o  et al. 2009;
Moore  et al. 2010; Raouafi et al. 2010; Hong et al. 2011).

\begin{figure}
   \centering
   \includegraphics[width=\textwidth, angle=0]{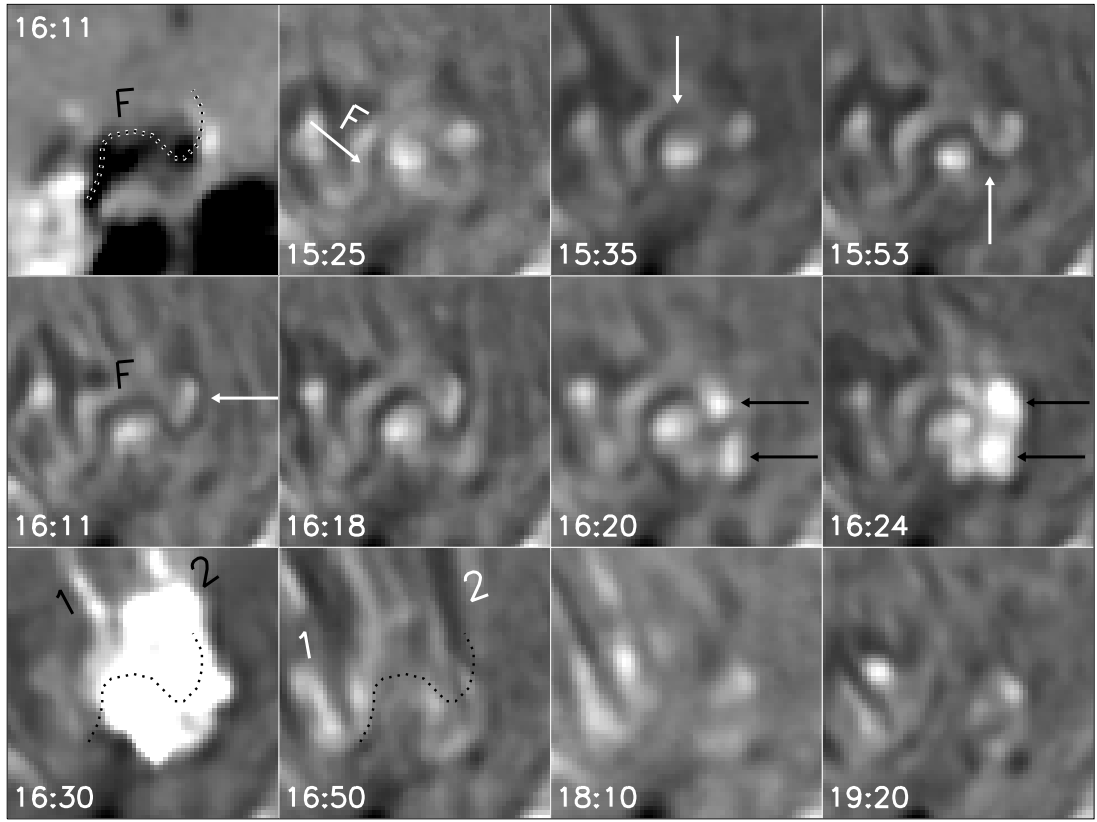}  
\caption{Close-up view of F in BBSO H$\alpha$ images. The first
panel is an MDI magnetogram. The white arrows indicate that F grew
toward the west, and the black arrows indicate the initial
brightenings of the flare showing two-ribbon nature. The outlines of
F's axis as in Fig. 1 are plotted, and its two legs are marked as
`1' and `2'. The FOV is $60^{''}$ $\times$ $60^{''}$. } \label{f2}
\end{figure}

\begin{figure}
  \centering
  \includegraphics[width=50mm, angle=0]{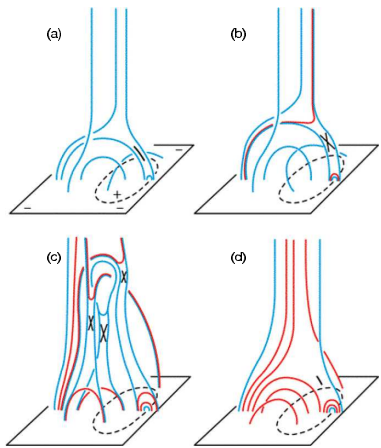}  
\caption{Schematic depiction of the production of a blowout coronal
jet from Moore  et al. (2010). Reproduced by permission of the AAS.
Only a few representative lines are drawn. Red field lines are those
that have been reconnected with reconnection-heated X-ray plasma on
them, while blue field lines either have not yet been reconnected or
will not be reconnected. ({\it a}) Pre-eruptive magnetic field
setup. A bipolar arch emerges in negative-polarity CH field. The
dashed oval is the neutral line around the positive flux of the
emerging arch, the short black curve represents the current sheet,
and the field line arching low along the neutral line in the middle
of the base arch indicates that the core field of the arch is
extremely sheared. ({\it b}) Magnetic reconnection (marked by the X)
at the location of the current sheet produces the small red loop on
the west side and the red field line anchored on the east side. If
the emerging arch has no sheared or twisted core field, the latter
forms a standard  X-ray jet and then the eruption would end shortly
after this time. ({\it c}) If the core field of the emerging arch is
strongly sheared and twisted, its blowout eruption leads to
reconnection at several locations (marked by the Xs) that produces a
blowout jet with a curtain-like structure, a red flare arcade in the
east of the neutral line, and in additional brightening and growth
of the loop in the west. ({\it d}) At the decay phase, the X-ray jet
narrows and the entire field structure begins to relax down to the
configuration of ({\it a}). } \label{f3}
\end{figure}

The schematic shown in Figure 3, a copy of Figure 10 in Moore
et al. (2010), depicts the scenario for the productions of both
standard and blowout jets. The magnetic field setup for both cases
is the same that a compact, low-arching bipolar field emerges into a
pre-existing, high-reaching, unipolar CH field with negative
polarity, thus forms a neutral line surrounding the positive base of
the emerging bipole, as well as a current sheet at the interface
between the ambient field and its positive-leg (Fig. 3a). If the
emerging-arch field has no appreciable shear or twist in it,
magnetic reconnection at the current sheet creates a
miniature-flare-arcade bight point over the west-side neutral line
below the reconnection X-point, as well as a standard X-ray jet with
a single-strand spire whipping out of the top side of the
reconnection X-point (Fig. 3b). The eruption process then finishes,
the field relaxes down to Fig. 3a configuration and the emerging
arch to the east of the new reconnection loop in Fig. 3b remains
stable. In the blowout jet case, however, an essential difference is
that the core field of the emerging bipolar arch is so strongly
sheared and twisted that it often shows up as a small filament and
has enough free energy to drive a blowout of the arch: the emerging
arch becomes destabilized by the earlier dynamics and thus triggers
a blowout eruption of the sheared core field as in a filament
eruption. As depicted in Fig. 3c, reconnection then occurs at three
places that can result in a blowout X-ray jet with a multi-stranded
curtain-like appearance (see Moore  et al. (2010) for details of the
blowout jet eruption process). Significantly, the blowout jet has a
substantially larger horizontal size span than that of the standard
jet. Finally, the jet narrows and the field configuration relaxes
down to the initial state of Fig. 3a (Fig. 3d). It is difficult to
make an one-to-one comparison between the above scenario and our
event because no soft X-ray or high-resolution EUV imaging
observations is available and the magnetic field environment is also
different from that described by Moore  et al. (2010) in coronal
holes. However, two factors make us believe that our H$\alpha$ surge
was clearly not a standard one. One is that it was originated from
the small F eruption, and the other is that its 1 and 2 components
quite resembled the curtain-like blowout jet rather than
single-strand spire of Moore et al. (2010). Therefore, we can
introduce a new term and call it ``blowout H$\alpha$ surge''. It is
interesting to note that the 1 and 2 features in our event were very
similar to a pair of thin ejections at the leg locations of a
spicule-associated blowout jet from an eruption of relatively cool
material in the event shown by Sterling, Harra, \& Moore (2010).
Taking 16:20 UT as the start time of the F eruption, F had a
lifetime of about 55 minutes from its first formation (15:25 UT)
through the eruption. This was well consistent with the 50-minute
mean lifetime of miniature erupting filaments given by Wang et al.
(2000). Similar to a case studied by Sakajiri et al. (2004), in
which a miniature filament erupted 25 minutes after its first
formation, we see that there was an interval of only 9 minutes
between the F eruption start and its full development (16:11 UT). It
is clear that the flare was preceded by such an eruption. Although
the flare at the main phase was too compact to discern its two
ribbons and separating motion (possibly due to the small size of F),
similar to the situation in compact flares observed by Tang (1985),
the small H$\alpha$ brightenings at the initial F eruption phase
clearly showed a two-ribbon nature. Consistent with some previous
observations (Wang et al. 2000; Sakajiri et al. 2004; Ren et al.
2008; Hong et al. 2011), therefore, we believe that the F eruption
was a small-scale version of active-region or quiet-region
large-scale filament eruptions and the flare was a direct result of
such a small-scale eruption.

\begin{figure}
   \centering
   \includegraphics[width=\textwidth, angle=0]{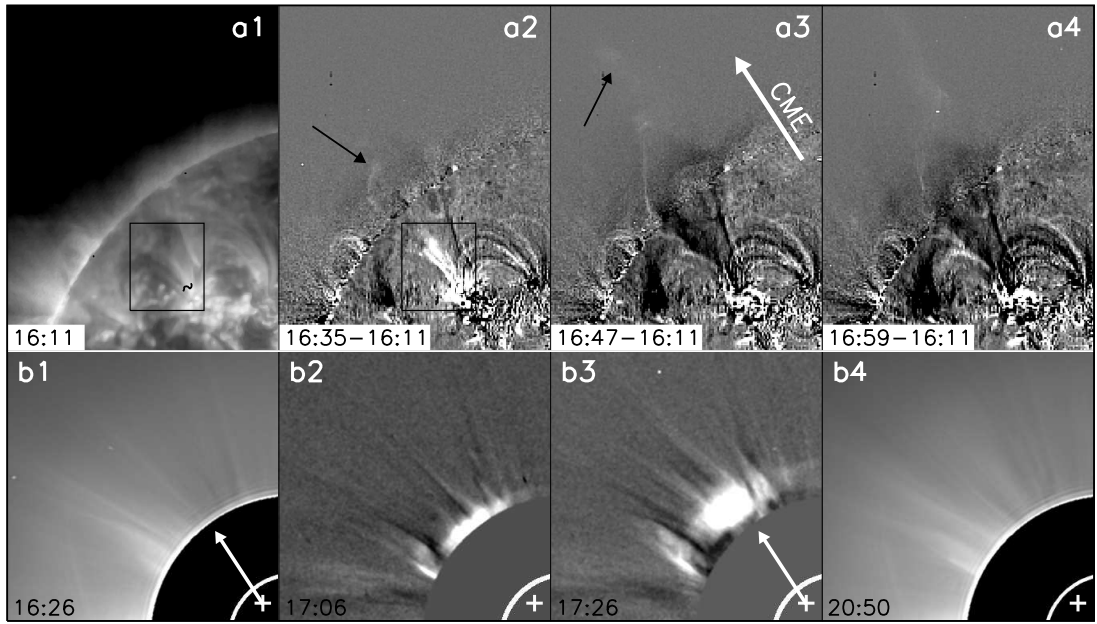}  
\caption{EIT 195 \AA\ (panels {\it a1-a4}) and LASCO C2 (panels {\it
b1-b4}) images. Panels {\it a1}, {\it b1}, and {\it b4} are original
images, {\it a2-a4} are EIT 195 \AA\ fixed-base difference images
obtained by subtracting a pre-flare image ({\it a1}), and {\it
b2-b3} are C2 running difference images. The black arrows indicate
the erupting F, and the white arrows the eruptive direction of the
CME. The outlines of F's axis as in Fig. 1 are also plotted in {\it
a1}, and the plus signs mark the pre-eruptive F location. The FOV
for {\it a1-a4} is $810^{''}$ $\times$ $1030^{''}$, and the black
boxes indicate the FOV of Fig. 1. } \label{f4}
\end{figure}

The eruption can be traced going beyond the northeast limb in EIT
observations, and was closely associated with a CME observed by
LASCO. This is quite obvious in the EIT 195 \AA\ and LASCO C2 images
presented in Fig. 4. At 16:35 UT, although there was a thick,
nearly straight, jet-shaped bright EUV feature responding to the
H$\alpha$ surge-like bright structure (see Fig. 1) on the solar disk
(included in the black box), the top part of the erupting F had
indeed ejected beyond the solar limb and showed a curved loop-like
shape (indicated by the black arrow). It can still be clearly
discernible at 16:47 UT but then sprayed out the field-of-view (FOV)
of EIT after 16:59 UT. Therefore, the EIT observation further
confirmed that this was a filament eruption event. The associated
CME first appeared in the LASCO C2 FOV at 17:06 UT as a ragged
ejection in the northeastern direction that traveled along a thin
streamer. It had a width of 66\degr\ and a central position angle
(P.A.) of 46\degr. As indicated by the white arrows, the CME's
eruption direction determined from the central P.A. was along the
streamer and consistent with that of the erupting F. It is
noteworthy that the streamer was only slightly altered after the
passage of the CME (see panels b1 and b4), which was similar to the
cases of streamer-puff and over-and-out CMEs recently studied by
Bemporad et al.(2005), Moore \& Sterling (2007), and Jiang et al. (2009). The
height-time (H-T) plot of the CME front at P.A. 38\degr\ is shown in
Fig. 5. By applying second-order polynomial fitting, back
extrapolation of the CME front from the H-T plots to the eruptive
location yields an estimate of the CME onset time near 16:17 UT,
which is very close to the start time of the F eruption at 16:20 UT,
as well as the start time of the flare at 16:25 UT. The spatial and
temporal consistencies strongly suggest that the CME initiation was
closely related to the F eruption. In addition, the application of
first- and second-order polynomial fitting to the H-T points gives
an average speed of 342 km $s^{-1}$ and an acceleration of -6.2 m
$s^{-2}$. These parameters showed that the CME was a slow one,
consistent with the results of that slow CMEs are more often found
associated with filament eruptions (Sheeley et al. 1999;
St. Cyr et al. 2000). It is
noted that the CME front speed is very close to the average F speed
of about 344 km $s^{-1}$ estimated from 16:20 to 16:47 UT, and two
H-T points of the erupting F at 16:35 and 16:47 UT plotted in Fig. 5
are also close to the extrapolated curves of the CME front, implying
the tight relationship between the F eruption and the CME. From
16:25 to 16:31 UT, however, we can exactly determine the top sites
of the erupting F in H$\alpha$ observations (see Fig. 1), and an
average speed (acceleration) of about 151.8 km $s^{-1}$ (1.4m
$s^{-2}$) is given. In accordance with the result of Zhang et al. (2001),
these values indicate that F possibly underwent a large acceleration
at its early eruption phase. As compared with the very small F size,
however, we would like to point out that the CME had much larger scale.
Therefore, the F eruption and the H$\alpha$ surge might only serve as a
trigger and occupy a small part of the CME. Further observations are
necessary to clarify such possibility.

\begin{figure}
   \centering
   \includegraphics[width=10cm, angle=0]{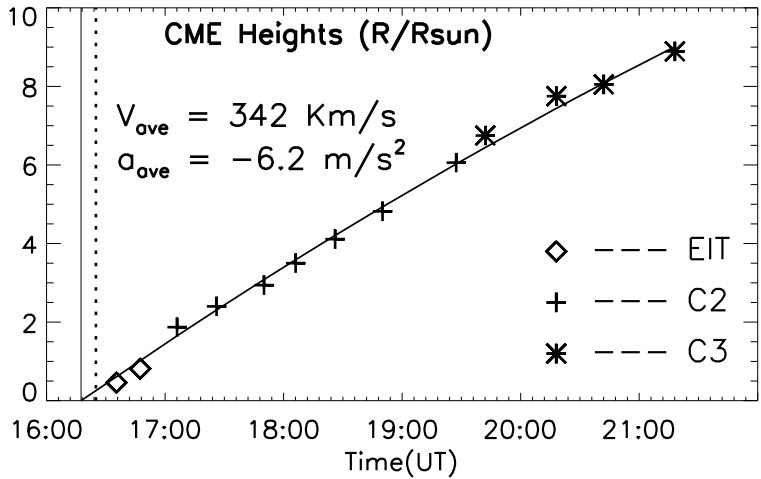}  
\caption{Heights of the CME front as the function of time, and the
back extrapolations by the use of second-order polynomial fitting.
The vertical dashed bar indicates the {\it GOES} flare start time,
and the solid bar the extrapolated CME onset time. } \label{f5}
\end{figure}

By further examining MDI observations, we found that the F eruption
was correlated with a distinct photospheric activity below it.
Fig. 6 shows the MDI magnetograms and continuum images around the
eruptive region. At first view, it is quite clear that the area of a
negative-polarity flux patch at the southern side of F greatly
decreased after the eruption (indicated by the white arrows). More
strikingly, a small pore just under the western part of F underwent
a sudden disappearance (indicated by the black arrows). The pore was
clearly seen at 15:59 UT before the eruption while it almost
completely disappeared in the next available MDI intensity image at
17:35 UT. Such disappearance thus took place in a timescale of less
than 100 minutes. Interestingly, it showed a $\delta$ configuration
by making a comparison between the 15:59 UT magnetogram and
continuum image. The flux changes in an area containing the eruptive
F (indicated by the white contours) and in a smaller area only
including the disappearing pore (indicated by the black dashed
boxes) are measured and plotted in Fig. 7, and the {\it GOES-8} 1-8
\AA\ soft X-ray flux profile is also overplotted to indicate the
flare time. After the flare start, we see that the negative flux in
the contour (box) area had an abrupt decrease on the order of 4.5
$\times$ $10^{19}$ Mx (1.7 $\times$ $10^{19}$Mx) within the flare
duration of only 7 minutes, i.e., impulsively dropped about 20\%
(34.7\%), which gives an average flux loss rate of about 1.1
$\times$ $10^{17}$ Mx$s^{-1}$ (4.0 $\times$ $10^{16}$ Mx$s^{-1}$).
In the following several hours, the change of the negative flux only
showed a simple tendency of continuous decrease, and had no
indication of a return to the pre-flare condition, indicating that
such an impulsive flux loss was permanent, not transient, and thus
was not due to the flare emission. We also note that the surrounding
positive flux only changed a little relative to the negative flux
during the flare. Therefore, the changes of opposite-polarity flux
in the contour (box) area were not simultaneous and seemingly not
balanced.

\begin{figure}
   \centering
   \includegraphics[width=\textwidth, angle=0]{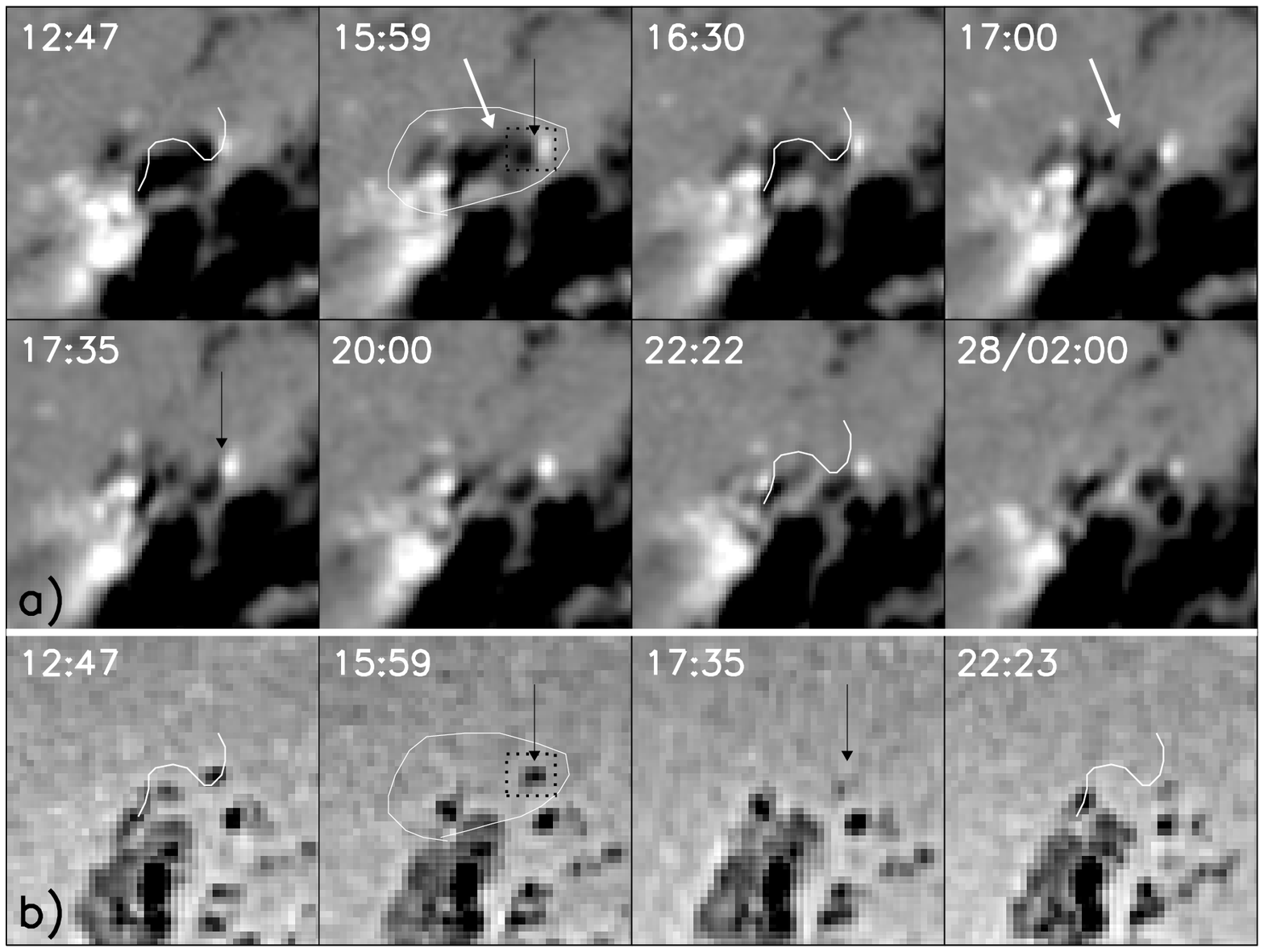}  
\caption{MDI magnetograms ({\it a}) and continuum images ({\it b})
around the eruptive filament. The white arrows indicate the negative
patch with a large reduction in area during the flare, and the black
arrows the vanishing small pore. The outlines of F's axis as in Fig.
1 are plotted, and the white contours and the black boxes indicate
two areas, in which the changes of magnetic flux are measured and
plotted in Fig. 6. The FOV is $90^{''}$ $\times$ $90^{''}$. }
\label{f6}
\end{figure}

\begin{figure}
   \centering
   \includegraphics[width=10cm, angle=0]{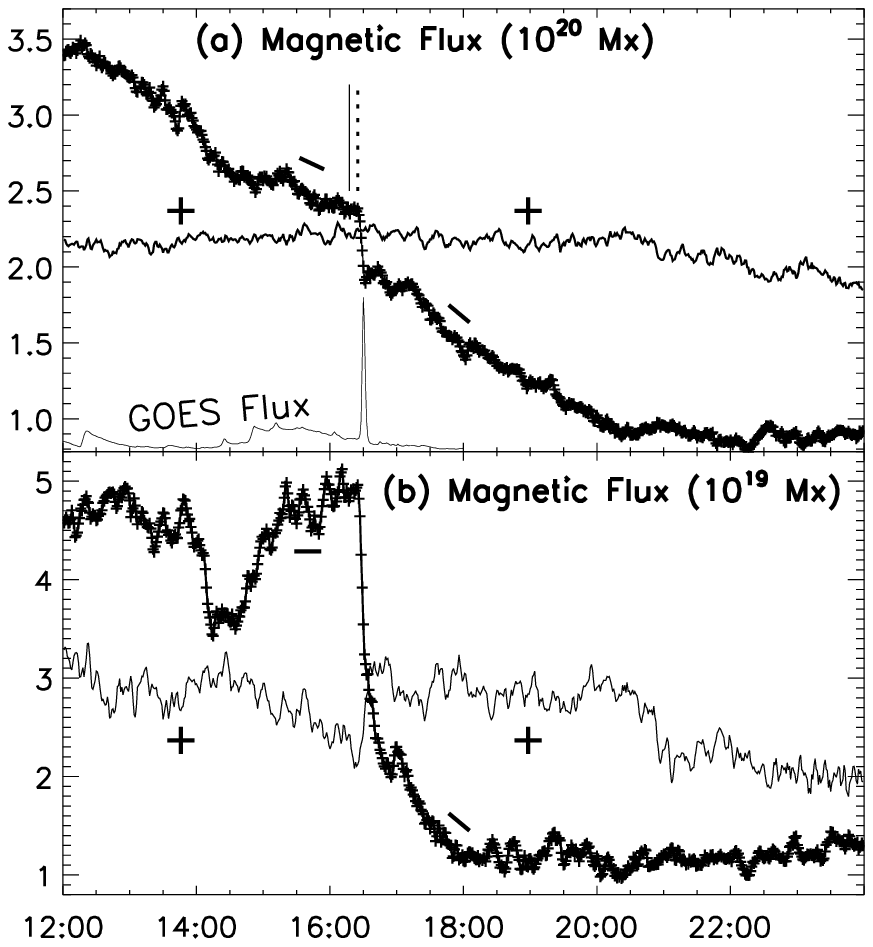}  
\caption{Changes of magnetic flux in the contour ({\it a}) and the
box ({\it b}) areas in Fig. 5, and time profiles of {\it GOES-8} 1-8
\AA\ soft X-ray, which are displayed in an arbitrary unit to fit in
the panel. The plus/minus sign marks the positive/negative flux. To
improve clarity, the absolute values for the negative flux are
plotted, and the positive flux values in panel {\it a} are shifted 1
on the vertical scale. The vertical dashed bar indicates the {\it
GOES} flare start time, and the solid bar the extrapolated CME onset
time. } \label{f7}
\end{figure}

\section{Conclusions and discussion}
\label{sect:Conclusion}
By means of high-resolution observations, the event was identified
as a surge-like eruption of the miniature filament F, i.e., a
blowout surge, instead of a standard surge according to the
following observations. (1) F divided opposite-polarity photospheric
magnetic field around the AR's northern periphery. (2) The eruption
was followed by a compact, impulsive flare. In the early eruption
phase, however, the initial flare brightenings had two ribbons. (3)
The evolution timescales of F, from its formation to full
development through eventual eruption, was similar to those in
previous studies. (4) The two legs of the erupting F were spatially
consistent with the two ends of the pre-eruption F. Our observations
further showed that the blowout surge was closely associated with
the CME along a preexisting streamer. Therefore, these observations
strongly indicated that the blowout surge, a filament eruption
taking a surge form, was a small-scale version of the large-scale
filament eruptions, with the exception of the small time and space
scales (Nistic\`o et al. 2009; Moore et al. 2010; Raouafi et al. 2010;
Hong et al. 2011).

An important aspect of the event is that, despite of the very small
size of F, its eruption and the following M2.2 flare were temporally
and spatially relevant to a common characteristic of photospheric
activity in major flares previously observed by some authors
(Kosovichev \& Zharkova 1999; Wang et al. 2002b; Meunier \& Kosovichev
2003; Li, Ding, \& Liu 2005; Liu et al. 2005; Sudol \& Harvey 2005;
Wang \& Liu 2010; Li et al. 2011), i.e., the sudden, significant, and persistent
decrease in negative flux on the order of 4.5 $\times$ $10^{19}$ Mx
and on a timescale of 7 minutes. As suggested by Sudol \& Harvey (2005),
similar changes are ubiquitous features of X-class flares and most
likely resulted from that the magnetic field changes direction
rather than strength, suggesting that they are consequences rather
than trigger of the flares. If so, the observed changes of magnetic
field in our example were probably due to that the field lines were
pulled or relaxed upward by the erupting F (also see Deng et al.
2005, Wang et al. 2005, and Hudson et al. 2008). As for the puzzling
signature of changes of flux imbalance in two polarities, although
some possible explanations were offered and discussed (Wang et al. 2002a;
Wang et al. 2002b; Liu et al. 2003), we also believe that it was mainly
associated with such changes in the magnetic field direction.
Another rare phenomenon in this example is that during the rapid
flux changes a small pore nearly disappeared in less than 100
minutes. Only one example of Wang et al. (2002a) has demonstrated that a
flare can be associated with complete disappearance of a small
sunspot in short period, thus our event gives another clear example
of that similar pore disappearance can also occur in a blowout
surge.

Another key question concerning the event is why F erupted in a
surge form? Since formation, maintenance, and instability of F
should be controlled by its magnetic field environment
(Martin 1990), we can tentatively consider the following three
factors. First, the very small size of the F is a main element,
which leads it to erupt in a narrow angular extent. We see that the
erupting F showed a large rising height relative to the width
between its two legs. Second, the eruption is strong enough to open
its overlying coronal arcade and escape out into the heliosphere,
which is indicated by the occurrence of the following flare and CME.
This is clearly different from the cases observed by Wang et al. (2000),
in which the most miniature erupting filaments are expelled to new
locations. Note that all the miniature filaments of Wang et al. (2000)
were on the quiet Sun with low magnetic flux density, while in our
example F was located in a region with relatively high flux density.
Therefore, it is naturally expected that a small change in the
stronger magnetic field might be enough to influence the topology
and stability of the F field in a violent manner and lead to a
strong eruption (Gaizauskas 1989). Finally, there exists a
guiding magnetic field with a twofold functions to channel the F
eruption and restrain its lateral swelling. The CME along the
streamer suggested that the F eruption from the streamer base should
be guided by its open field component. This is very similar to that
occurring in standard surges, which could also be ejected along
either open magnetic fields or closed large loops. It is noteworthy
that, however, the physical scenarios of standard and blowout surges
are different. As shown in Fig. 3, magnetic reconnection only occurs
between the guiding field and the closed magnetic field of new emerging
flux in the standard surge case, while in the blowout surge case
reconnection might occur at different places (Moore et al. 2010).
Therefore, a clear distinction between standard and blowout surges
by using of high-resolution observations is necessary to understand
their physical natures and the origin of associated eruptive
phenomena. We can imagine that if we only have H$\alpha$
observations with poorer spatial resolution, the blowout surge in
this example will be readily regarded as a standard one, and thus
might lead to misunderstand the origin of the associated CME. Even
though the spatial resolution of our observations is high enough,
the two phenomena are so easily mistakable for each other that we
should differentiate them with great care. It is anticipated that
observations from the {\it Solar Dynamics Observatory} (SDO) would
be great beneficial in this problem.

\begin{acknowledgements}
We thank an anonymous referee for many constructive suggestions
and thoughtful comments that helped to improve the clarity and
quality of this paper.
We are grateful to the observing staff at BBSO for making good
observations. We thank the {\it SOHO}/MDI, EIT and LASCO teams for
data support. This work is supported by the 973 Program
(2011CB811400) and by the Natural Science Foundation of China under
grants 10973038 and 11173058.
\end{acknowledgements}

\label{lastpage}

\end{document}